\documentclass[preprint,preprintnumbers,showpacs,amsmath,amssymb]{revtex4}
\usepackage{graphicx}
\usepackage{dcolumn}
\usepackage{bm}
\usepackage[usenames]{color}
\begin{document}
\title{A note on the FKPP equation approached with the hyperbolic scaling}
\author{M.A. Reyes}
\email{marco@fisica.ugto.mx}
\affiliation{Departamento de F\'{\i}sica, DCI Campus Le\'on, Universidad de Guanajuato, Apdo. Postal E143, 37150 Le\'on, Gto., Mexico.}
\author{H. C.  Rosu}
\email{hcr@ipicyt.edu.mx}
\affiliation{IPICYT, Instituto Potosino de Investigacion Cientifica y Tecnologica,\\Apdo Postal 3-74 Tangamanga, 78231 San Luis Potos\'{\i}, S.L.P., Mexico.}
\date{\today}
\begin{abstract}
We consider the hyperbolic scaling of the FKPP equation and introduce two solutions of the action functional type in the limit of zero hyperbolic parameter. Furthermore, we show that the action functional of the Ablowitz-Zepetella kink is a special case of one of those solutions.
\\

\end{abstract}
\pacs{\\
{\tt 87.10.Ed} -- Ordinary differential equations (ODE), partial differential equations (PDE), integrodifferential models \\
{\tt 82.40.-g} -- Chemical kinetics and reactions: special regimes and techniques\\
{\tt 87.23.Cc} -- Population dynamics and ecological pattern formation
%
}

\maketitle


Parabolic and hyperbolic transformations of the independent variables are well-known mathematical features of the partial differential equations.
In the case of the heat equation $u_t=\nabla^2u$ for example, the parabolically-scaled solution $u(\gamma x, \gamma^2 x)$ also solves the equation for any positive $\gamma$ parameter. In other words, the standard solution $u(x,t)$ is the special case $\gamma=1$ of the parabolically-scaled family of solutions describing the long-time large-distance asymptotics of more microscopic transport phenomena. As pointed out by Fedotov \cite{F2000}, the propagating fronts of reaction diffusion equations should be approached with the more `democratic' hyperbolic scaling $t \rightarrow \frac{t}{\epsilon}$, $x \rightarrow \frac{x}{\epsilon}$ ($\epsilon \ll 1$) if one wants to include their microscopic origin.
Following Fedotov, we consider the Fisher-Kolmogorov-Petrovskii-Piskunov equation for a scalar field $\rho(x,t)$
\cite{F2000}
\begin{equation}\label{fk1}
\frac{d\rho}{dt}=D \frac{d^2\rho}{dx^2}+U\rho(1-\rho)~,
\end{equation}
where $D$ is the diffusion constant and $U$ is the constant reaction rate. The logistic term describes the spreading of the scalar field in an environment, originally applied to the spreading of an advantageous gene in a population. Using the hyperbolically-rescaled field
\begin{equation}\label{fk2}
\rho\left(\frac{x}{\epsilon},\frac{t}{\epsilon}\right)=\rho^{\epsilon}(x,t)~,
\end{equation}
one gets
\begin{equation}\label{fk3}
\epsilon\frac{d\rho^\epsilon}{dt}=\epsilon^2 D \frac{d^2\rho^{\epsilon}}{dx^2}+U\rho^{\epsilon}(1-\rho^{\epsilon})~.
\end{equation}
We also notice that if in this scaled equation we substitute the imaginary scaling parameter $\epsilon=i\hbar$, we obtain formally a nonlinear cubic Schr\"odinger equation,
where $D$ is identified with the inverse of the mass of a quantum entity, whereas the reaction parameter is now the strength of the nonlinear interaction. However, this is not our concern in the following.

If we make the change of dependent variable $\rho^\epsilon=e^{-\frac{G(x,t)}{\epsilon}}$ the following equation is obtained
\begin{equation}\label{fk4}
\frac{\partial G}{\partial t}=\epsilon D \frac{\partial ^2G}{\partial x^2}-D\left(\frac{\partial G}{\partial x}\right)^2-U\left(1-e^{-\frac{G}{\epsilon}}\right)~.
\end{equation}
In addition, in the limit $\epsilon\rightarrow 0$ the above equation reduces to:
\begin{equation}\label{fk5}
\frac{\partial G}{\partial t}+ D \left(\frac{\partial G}{\partial x}\right)^2+U=0~.
\end{equation}
The latter equation is the Hamilton-Jacobi equation $\frac{\partial G}{\partial t}+H\left(\frac{\partial G}{\partial x}\right)=0$ for the Hamiltonian $H(p)=Dp^2+U$, where $G$ plays the role of the action functional. To solve this Hamilton-Jacobi equation we propose the ansatz $G(x,t)=cx^at^b-\alpha t$, which gives
\begin{equation}\label{fk6}
bcx^at^{b-1}-\alpha+Dc^2a^2x^{2(a-1)}t^{2b}+U=0~.
\end{equation}
Thus, $\alpha=U$, $a=2$, and $b=-1$, which provide $c=1/4D$ and the solution:
\begin{equation}\label{fk7}
G_1(x,t)=\frac{x^2}{4Dt}-Ut~.
\end{equation}
The reaction front can be found from the conditions $G(t,x(t))=0$ and $x=vt$, implying that the reaction velocity is $v=\sqrt{4DU}$. Solution (\ref{fk7}) is also given by Fedotov but here we obtained it by other means.

We here write down another solution, which comes out from the similarity with the harmonic oscillator solution of the HJ equation $G_{osc}=W(x)-\beta t$:
\begin{equation}\label{fk8}
G_2(x,t)=\sqrt{\frac{\beta-U}{D}} \left(x-\frac{\beta}{\sqrt{\frac{\beta-U}{D}}}t\right)~.
\end{equation}
This solution has not been given in the literature, although it is easily obtainable through the HJ route. Indeed, since $\frac{\partial G}{\partial x}=\frac{dW}{dx}$, equation (\ref{fk5}) reduces to the following form
\begin{equation}\label{fk8bis}
-\beta+D\left(\frac{dW}{dx}\right)^2+U=0~,
\end{equation}
which immediately leads to the solution (\ref{fk8}) if the initial condition $x(t=0)=0$ is used for the propagation front.

Finally, a third form of the solution can be obtained if it is asked to be directly of traveling type, i.e., $G(x,t)=G_3(z)$, $z=x-vt$. Equation (\ref{fk5}) takes the form
\begin{equation}\label{fk9}
D\left(\frac{d G_3}{dz}\right)^2-v\frac{dG_3}{dz}+U=0
\end{equation}
and the solution will be
\begin{equation}\label{fk10}
G_3(x,t)=\frac{v\pm\sqrt{v^2-4DU}}{2D}(x-vt)~.
\end{equation}
One can see that if $v^2-4DU=0$, then $p=\frac{\sqrt{4DU}}{2D}$, thus again the mass of the equivalent microscopic entity is $m=\frac{1}{2D}$.

We want also to discuss in this context the well-known Ablowitz-Zeppetella solution of the FKKP equation \cite{AZ79}, following the results of a paper by Rosu and Cornejo-P\'erez \cite{RCP05}, who obtained it through a factorization of equation (\ref{fk3}). Writing the equation in the form
\begin{equation}\label{rc1}
\frac{\partial \rho^\epsilon}{\partial t}=\epsilon D \frac{\partial^2\rho^{\epsilon}}{\partial x^2}+\frac{U}{\epsilon}\rho^{\epsilon}(1-\rho^{\epsilon})~,
\end{equation}
introducing the new variables $\tilde{t}=\frac{U}{\epsilon}t$ and $\tilde{x}=\frac{1}{\epsilon}\left(\frac{U}{D}\right)^{1/2}x$ and passing to the traveling coordinate $\tilde{z}=\tilde{x}-\tilde{v}\tilde{t}$, one gets the ordinary differential equation
\begin{equation}\label{rc2}
\rho''+\tilde{v}\rho'+\rho(1-\rho)=0~,
\end{equation}
where the prime denotes the derivative $d/d\tilde{z}$. According to \cite{RCP05}, for $\tilde{v}=\frac{5}{\sqrt{6}}$ equation (\ref{rc2}) can be factorized in the way they proposed which leads to a first order differential equation easy to integrate. The solution is precisely the Ablowitz-Zepetella solution, i.e.,
\begin{equation}\label{rc3}
\rho(\tilde{z})=\frac{1}{4}\left(1-\tanh \frac{\tilde{z}}{\sqrt{24}}\right)^{2}~.
\end{equation}
In the asymptotic limit $\tilde{z}\rightarrow \infty$, we get $\rho \sim e^{-\sqrt{\frac{2}{3}}\tilde{z}}$. This leads to the following action functional
\begin{equation}\label{rc4}
G_{AZ}(x,t)=\sqrt{\frac{2U}{3D}} \left(x-\sqrt{DU}t\right)~.
\end{equation}
It is worth noting that $G_{AZ}$ is identical to $G_2$ in the special case $\beta=\sqrt{\frac{2}{3}}U$. The latter value implies a negative reaction rate, i.e., a rate of disappearance of the reactant.

In conclusion, we obtained two novel solutions of traveling type of the Hamilton-Jacobi equation corresponding to the FKPP equation in the limit of zero hyperbolic parameter. In addition, we have shown that the action functional of the Ablowitz-Zepetella kink solution of the FKPP equation corresponds to a special case of one of those solutions.

\end{document}